\documentclass[aps, prl, reprint, superscriptaddress, showpacs]{revtex4-1}
\usepackage{graphicx, color}
\usepackage{amsmath}
\usepackage{subsupscripts,epstopdf}
\usepackage[disable]{todonotes}

\begin{document}

%

\title{Orientation Resolution through Rotational Coherence Spectroscopy}



\author{Varun Makhija}
\affiliation{J. R. Macdonald Laboratory, Kansas State University, Manhattan, KS 66506}
\affiliation{University of Ottawa, Department of Physics, MacDonald Hall 150 Louis Pasteur Ottawa ON Canada K1N 6N5}
\author{Xiaoming Ren} 
\altaffiliation{Current address: Institute for the Frontier of Attosecond Science and Technology, CREOL and Department of Physics, University of Central Florida, Orlando, Florida 32816, USA.}
\author{Drue Gockel}
\author{Anh-Thu Le} 
\author{Vinod Kumarappan}
\email[Email: ] {vinod@phys.ksu.edu}


\date{\today}

\begin{abstract}
The rich information content of measurements in the molecular frame rather than the laboratory frame has motivated the development of several methods for aligning gas phase molecules in space. Even so, for asymmetric tops the problem of making molecular frame measurements remains challenging due to its inherently multi-dimensional nature. In this Letter we present a method, based on the analysis of observables measured from rotational wavepackets, that does not require either 3D alignment or coincident momentum measurements to access the molecular frame. As an application we describe the first fully-orientation-resolved measurements of strong-field ionization and dissociation of an asymmetric top (ethylene). 
\end{abstract}
\pacs{37.10.Vz, 33.15.Bh, 33.80.-b,33.20.Sn,33.80.Rv,33.80.Wz}
\maketitle

Our understanding of gas-phase photochemical processes has long been impeded by the necessity of averaging over the distribution of molecular orientations encountered in the laboratory. This averaging often severely degrades the quality of information and insight that can be obtained from an experiment. For instance, the critical role that multiple orbitals play in high harmonic generation from molecules went under-appreciated for many years until molecules with a narrow orientation distribution generated by intense-laser alignment~\cite{friedrich1995,kim1996,sakai1999,tamar1999,roscapruna2001}, were used~\cite{mcfarland2008,smirnova2009}. Hexapole focusing combined with brute-force electric field orientation~\cite{kramer1965} was similarly instrumental in stereochemistry~\cite{bernstein1987}, facilitating the measurement of orientation dependent reaction and ionization rates ~\cite{brooks1966,kaesdorf1985}. The development of powerful multi-particle coincidence techniques~\cite{eland1972,eland1979}, which rely on identifying the orientation of each molecule from the momenta of charged fragments, has enabled copious molecular frame measurements such as the determination of molecular frame photoelectron angular distributions of linear ~\cite{golovin1990,shigemasa1995} and symmetric top molecules~\cite{williams2012}. However, due to geometric restrictions in alignment experiments the entire space of orientations cannot be accessed, and molecular frame measurements with the coincidence method can only be made for dissociative processes for which the axial recoil approximation holds.

In this letter we present a complementary route to the molecular frame that overcomes these limitations. A new class of experiments, which relies neither on preselecting particular orientations of molecules by aligning them nor on post-selection by sorting the molecules by their experimentally determined orientation but on the coherent evolution of a rotational wavepacket, has recently emerged.  By careful analysis of variation in the effects of a probe laser pulse as the rotational wavepacket launched by an ultrashort pump pulse evolves,  the molecular frame angular dependence of the probe process can be extracted.  But this type of analysis has so far been limited either to linear molecules \cite{kanai2005,thomann2008,vozzi2011,weber2013,ren2013}, approximately symmetric top molecules \cite{mikosch2013a,mikosch2013b}, or to the lowest order for all axes of an asymmetric top \cite{spector2014}. We generalize this method to enable the extraction of the complete angle dependence of a light-induced process in an asymmetric top molecule. We apply the method to the strong-field ionization and dissociation of ethylene in a linearly polarized femtosecond laser pulse. In the process, we also characterize the pump-driven rotational wavepacket in two Euler angles. The method, which extracts Orientation Resolution from Rotational Coherence Spectroscopy (ORRCS) is expected to be applicable to a wide variety of light-initiated processes in asymmetric top molecules.

\begin{figure}
    \includegraphics[width=1\columnwidth]{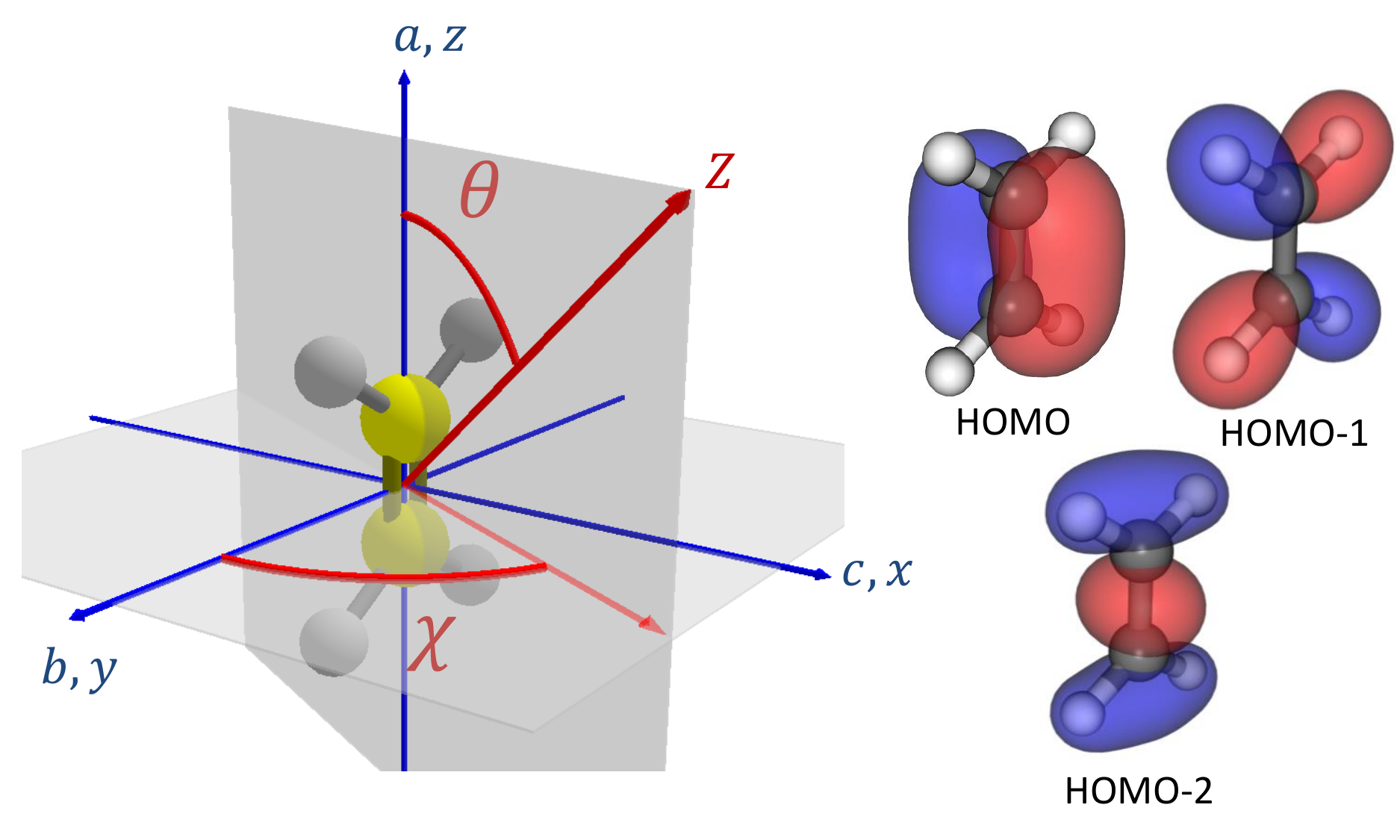}
    \caption{In the coordinate system used the molecular frame is defined by the principal axes of rotation $a$,$b$ and $c$ show in the figure. The Euler angles $\theta$ and $\chi$ are the polar and azimuthal angles of the lab frame $Z$ axis defined the laser polarization. The maximum amplitude isosurfaces of the HOMO (I$_p=$10.51~eV), HOMO-1 (I$_p=$12.82~eV) and HOMO-2 (I$_p=$14.96~eV) orbitals are shown below with the ionization potentials for each orbital given.}
    \label{fig:euler}
\end{figure}
\begin{figure*}
    \includegraphics[width=2\columnwidth]{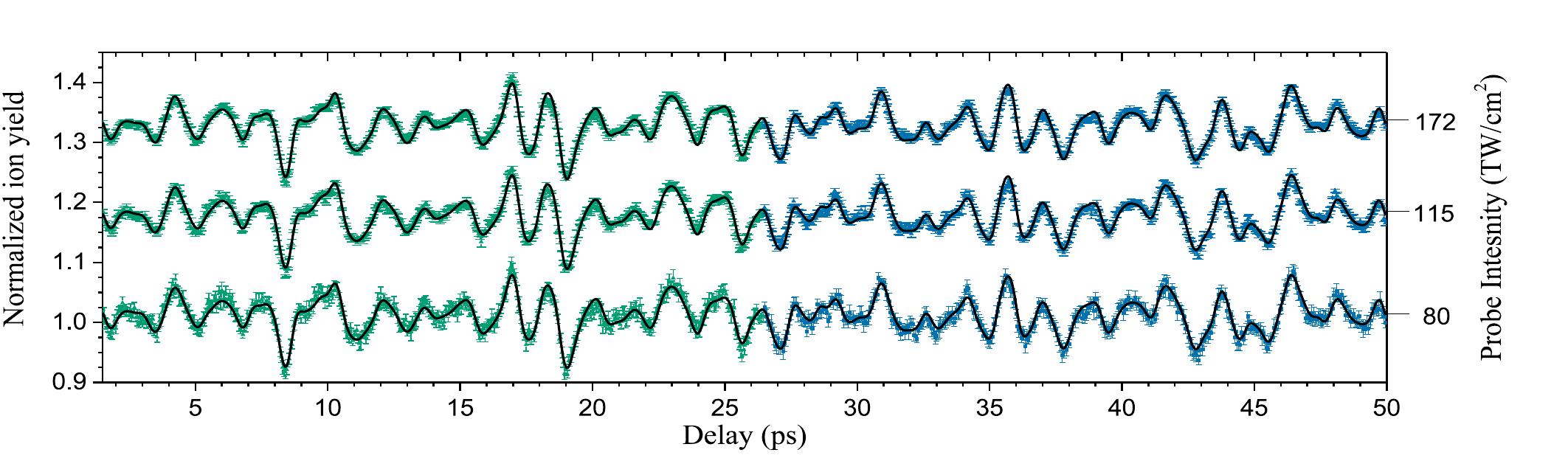}
    \caption{The normalized yield of C$_2$H$_4^+$ with three different probe intensities and the same pump intensity. Data points shown in green (up to 26.45~ps) are used for the fit. The black line is the resulting fit function calculated over the entire window. Data shown in blue (after 26.45~ps)were not used in the fit and are well modeled by the fit function.}
    \label{fig:all_data}
\end{figure*}
For an asymmetric top, the orientation dependence of any process driven by a linearly polarized laser pulse can be expanded in a basis of matrix elements of the infinitesimal rotation operator - $D_{mk}^j(\phi,\theta,\chi)$ - which form a complete basis for functions on the rotation group SO(3)~\cite{wigner},    
\begin{equation}
S(\theta,\chi) = \sum_{jk} C_{jk} D_{0k}^j(\phi,\theta,\chi).
\label{eq:model}
\end{equation}
Since we are considering only linearly polarized light, the function $S$ is independent of the azimuthal Euler angle $\phi$; hence $m=0$. In the molecular frame the Euler angles $\theta$ and $\chi$ are the polar and azimuthal angles of the laser polarization vector, respectively as shown in Fig.~\ref{fig:euler}. Symmetry considerations that further restrict this sum in the particular case of ethylene are explained in the supplemental material (SM). If the measurement is made not in the molecular frame but from a rotational wavepacket excited by a preceding pump pulse, the delay-dependent expectation value of the angle-integrated yield
\begin{eqnarray}
\left<S\right>(t) &=& \int\rho(\theta,\chi,t)S(\theta,\chi)\sin\theta d\theta d\chi \nonumber\\
&=&\sum_{jk} C_{jk} \int\rho(\theta,\chi,t)D_{0k}^j(\phi,\theta,\chi)\sin\theta d\theta d\chi \nonumber\\
&=&\sum_{jk} C_{jk}\left<D_{0k}^j\right>(t).
\label{eq:fit}
\end{eqnarray}
Here, $\rho(\theta,\chi, t)$ is the delay-dependent molecular axis distribution. The problem of finding $S(\theta,\chi)$ is thus reduced to determining the coefficients $C_{jk}$ from the expansion of the measured time-dependent signal $\left<S\right>(t)$ in terms of $\left<D_{0k}^j\right>$. This assumes that the $\left<D_{0k}^j\right>$ are known (or, equivalently, the rotational wavepacket is known). We will show that the full axis distribution $\rho(\theta,\chi,t)$ can be determined from the data itself via comparison with the $\left<D_{0k}^j\right>$ calculated from the rigid-rotor time-dependent Schr\"{o}dinger equation~\cite{tamar1999,rouzee2008,pabst2010,makhija2012}, and that the coefficients can be obtained when the series is truncated appropriately. We show that using a long high-resolution delay scan obviates the need for direct angle dependent measurements and makes available 2D angular information from asymmetric top molecules.
\begin{figure*}
    \centering
        \includegraphics[width=2\columnwidth]{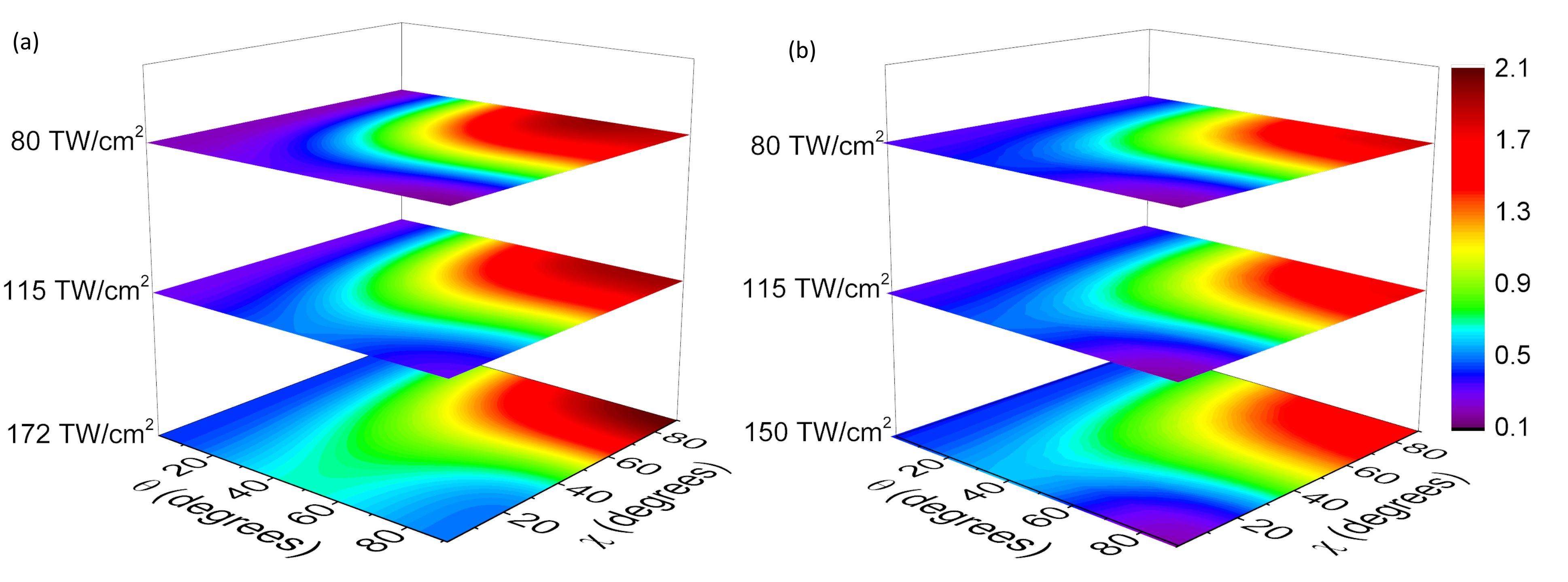}
    \caption{(a) Experimentally determined and (b) theoretically calculated  molecular frame, angle dependent C$_2$H$_4^+$ yields.From the angular dependence the ionizing orbital can determined (see text)}
    \label{fig:ion2D}
\end{figure*}

In the first experiment, we launch a rotational wavepacket in ethylene (C$_2$H$_4$) molecules, cooled by supersonic expansion from a high pressure jet~\cite{even2000} (20\% ethylene and balance helium at a total pressure of 70 bar) and skimmed into a velocity map imaging spectrometer \cite{eppink1997}, with a 4 TW/cm$^2$, 120~fs, pump pulse. The molecules are then ionized with a 30~fs probe pulse that is delayed with respect to the pump pulse by up to 50~ps. At the highest probe intensity used in the experiment $\leq5\%$ fragmentation is observed in the time-of-flight spectrum. The yield of C$_2$H$_4^+$ ,a measure of the effectiveness of ionization of the molecule in the probe pulse, is recorded as a function of pump-probe delay by setting the gate of a boxcar integrator on the molecular ion peak in the time-of-flight spectrum. Note that momentum information is not recorded in this experiment. By using an optical chopper and the acquisition scheme used previously for optical measurements of alignment \cite{ren2012}, the ionization rate from unaligned ethylene and from background gas is also recorded at the same time. The yield from the aligned molecules is then normalized to the yield from unaligned molecules after background has been subtracted from both. To reduce the effect of drifts in laser and jet conditions $10$ scans are performed depending on the contrast of the TOF peak and then averaged. The data are shown in Fig.~\ref{fig:all_data}; the error bars show the statistical standard deviation and the data for each probe intensity have been offset for display.

Only the first 500 (up to 26.45~ps) data points shown in green are used to extract the coefficients $C_{jk}$ as well as the molecular axis distribution seen by the probe pulse. Note that the molecular axis distribution is an implicit and non-linear function of pump intensity and pulse duration, and of the rotational temperature of the gas. But, for a given set of values for these three parameters, Eq.~\ref{eq:fit} is linear in the $C_{jk}$'s. Hence we use linear regression to determine the $C_{jk}$'s for a table of laser parameters and gas temperatures (cf. SM). For this purpose, we first solve the TDSE for  34 laser intensities (1.0 to 12.8 TW/cm$^2$ in steps of 0.2 TW/cm$^2$), 14 pulse durations (60 to 200 fs in steps of 10 fs) and 175 initial rotational states (these suffice to construct thermal distributions for any temperature below 15~K) under the assumption that the pump pulse does not excite any vibrational or electronic states (the rigid rotor approximation) \cite{rouzee2008}. The $\left<D_{0k}^j\right>$ functions up to $j=8$, $k=8$ for each rotational state are calculated as a function of delay (up to 100 ps) and stored. For each pulse duration, intensity and temperature we determine the values of C$_{jk}$  that minimizes the squared difference between the measured and computed signal by linear regression. The sum is terminated based on the variability of the coefficients as a function of the number of data points included in the fit (cf. SM). The resulting error surface converges to a curve of constant fluence for the minimum reduced $\chi^2$, indicating that the alignment is truly impulsive~\cite{lotte2007}. In Figure~\ref{fig:all_data}, the data and the best fit curves for C$_2$H$_4^+$ yield are shown superimposed on the data for three different probe laser intensities. While the fits are performed independently the retrieved rotational temperature is 9~K in each case. The best fit pump fluence values of 456, 520 and 492~mJ/cm$^2$ for the 80, 115 and 172 TW/cm$^2$ probe pulses deviate at most by $9.7\%$ from the measured value of 480~mJ/cm$^2$. We would like to note here that in cases where intensity averaging plays a significant role, the best fit laser parameters may not coincide with those measured in the lab. We try to minimize the effect of intensity averaging by expanding the 1~cm diameter laser beam to $\approx$ 1.5~cm in the probe arm and shrinking it to $\approx$ 0.6~cm in the pump arm before focusing. In all cases the fit functions $S(t)$ plotted in black over the entire delay window in Fig.~\ref{fig:all_data} accurately model the blue data points (after 26.45~ps) not included in the fit confirming the reliability of the extracted coefficients.

\begin{figure*}
    \centering
        \includegraphics[width=2\columnwidth]{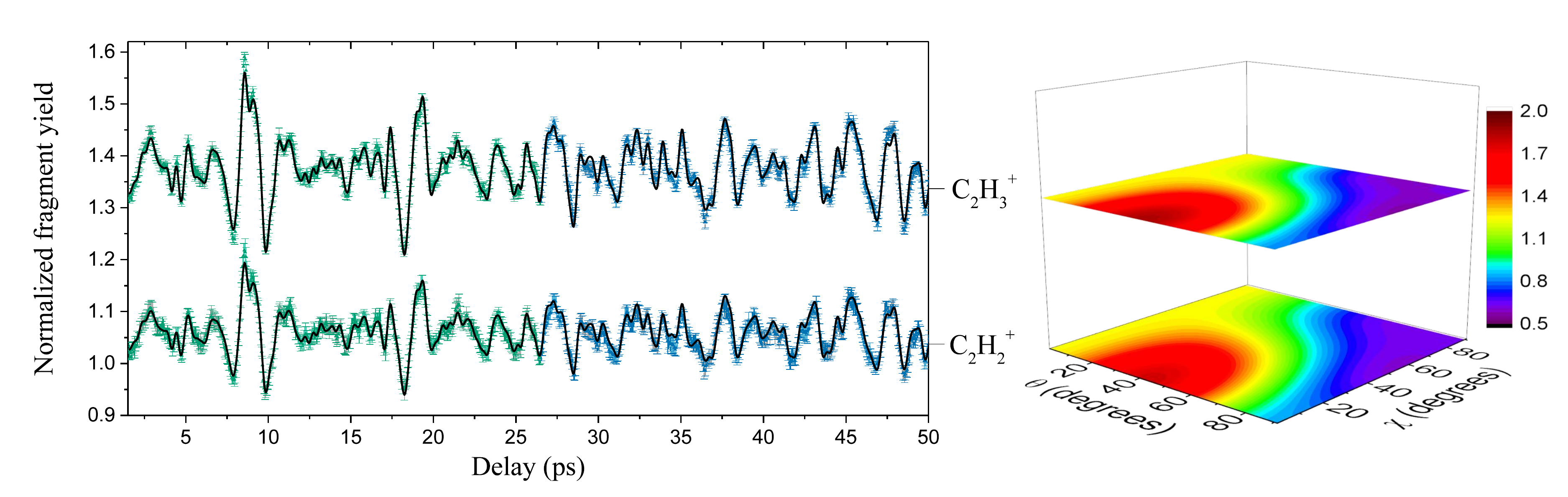}
    \caption{Left: Measured delay dependent yields of C$_2$H$_2^+$ and C$_2$H$_3^+$. Data points shown in green (up to 26.45~ps) are used for the fit. The black line is the resulting fit function calculated over the entire window. Right: Resulting molecular frame, angle dependent fragmentation rates of ethylene. From the shape of the angle dependence it can be concluded that fragmentation initiated by the strong field proceed via removal of HOMO-1 or HOMO-2 electrons (see text).  }
    \label{fig:frag2D}
\end{figure*}
The 2D angle dependent ionization yields extracted from the from the data in Fig.~\ref{fig:all_data} are shown in Fig.~\ref{fig:ion2D} (a). These are normalized to the yield from an isotropic sample. This measurement of the 2D angular dependence of non-dissociative ionization illustrates the power of ORRCS. At all three probe intensities the ionization probability has a maximum at $\theta$ and $\chi=90^{\circ}$, consistent with the density profile of the highest occupied molecular orbital (HOMO), shown in Fig.~\ref{fig:euler} with HOMO-1 and HOMO-2. At $\chi=0^{\circ}$, where the HOMO has a node, we see a modulation which peaks near $\theta=45^{\circ}$ consistent with the density profile of HOMO-1. From these observations we can conclude that the ground and first exited states of the molecular ion are populated. Fig.~\ref{fig:ion2D} (b) shows calculations in the strong field approximation of normalized ionization yields from HOMO, HOMO-1 and HOMO-2. As the intensity of the probe is increased the strength of the modulation at $\chi=0^{\circ}$ gets amplified, indicating that the contribution from HOMO-1 relative to that from HOMO increases with intensity and is responsible for the variation of the angle dependence. It was also necessary to include HOMO-2 in the calculations to account for the non-zero yield at the nodes of the HOMO and HOMO-1 orbitals \cite{lotte2010}, although its integrated contribution is only 0.0005\% of the total, integrated yield.

In the second experiment, we increase the intensity of the probe pulse until fragment peaks are observed in the time-of-flight mass spectrum. We then measure delay dependent yields for each channel from rotationally excited molecules by moving the gate of the boxcar integrator from one peak to the next. The data were collected at an ionizing intensity of about 250 TW/cm$^2$ where no charged carbon fragments are detected. Time dependent yields for C$_2$H$_2^+$ and C$_2$H$_3^+$ are shown in Fig.~\ref{fig:frag2D} with only the data points shown in green used for the fit. The retrieved temparature is 7~K in each case. The extracted fluence values of 1680 and 1716 mJ/cm$^2$ deviate by $6.8$ and $4.8\%$ respectively from the measured value of 1802 mJ/cm$^2$. Also shown in Fig.~\ref{fig:frag2D} are the angle dependent yields for C$_2$H$_2^+$ and C$_2$H$_3^+$ fragments. The previously measured appearance energies of these fragments are consistent with population of the ionic $\tilde{A}$ $\lsuperscript{B_{3g}}{2}$ \footnote{The states will be labeled with the subscripts g or u indicating even or odd inversion symmetry as they are in the literature. We note however that the ground state of the ion has a twisted stable geometry, thereby lowering its symmetry to lowering to $\lsuperscript{B_{3}}{2}$} (removal of a HOMO-1 electron) or $\tilde{B}$ $\lsuperscript{A_{g}}{2}$ (removal of a HOMO-2 electron) states \cite{lorquet1980,kim2005}; indicating that population of these states leads to the ejection of H and H$_2$. In our angle dependent data we observe that the fragmentation process preferentially selects molecules aligned near $\theta=45^{\circ}$ and $\chi=0^{\circ}$ consistent with fragmentation following ionization into the $\tilde{A}$ $\lsuperscript{B_{3g}}{2}$ state. The yield is also enhanced at $\theta=0^{\circ}$ for any value of $\chi$, consistent with fragmentation initiating from the $\tilde{B}$ $\lsuperscript{A_{g}}{2}$ state. We can thus conclude that fragmentation initiated by the intense probe proceeds via ionization into both these states~\cite{xie2014}. However at intensities below $\sim$ 180 TW/cm$^2$ we observe evidence of dissociative ionization into the $\tilde{A}$ $\lsuperscript{B_{3g}}{2}$ state (cf. fig.~\ref{fig:ion2D}). This may be explained by the fact that the removal of H/H$_2$ requires 0.68/0.75 eV additional energy over the ionization threshold into the $\tilde{A}$ $\lsuperscript{B_{3g}}{2}$ state and depends on the vibrational modes excited in this state \cite{stockbauer1975,lorquet1980,kim2005,baptiste2014}. 

Since our analysis extracts the best fit values for fluence as well as the rotational temperature of the gas, it effectively determines the time-dependent molecular axis distribution as well. The experimental determination of this distribution has heretofore been plagued with difficulties. Coulomb explosion imaging measurements would require not only the coincident measurement of at least three fragments, but also prior knowledge of the orientation selectivity of the probe~\cite{lee2006,williams2012,underwood2015}. Optical measurements, for which the relevant non-linear optical tensor may be known, provide too little information to directly infer even the expectation vales of the lowest order moments of the molecular axis distribution~\cite{rouzee2008}. But our analysis could be applied to either measurement to fully characterize both the wavepacket and the probe process.  
 
Rotational coherence spectroscopy, a technique pioneered by Zewail and Felker~\cite{felker1992}, relied on the knowledge of $S(\theta,\chi)$ for one- and two-photon processes to obtain the moment of inertia tensor---information about the static geometry of molecules. By approaching rotational wavepacket dynamics from a different perspective, in which the molecular structure is assumed to be known but $S(\theta,\chi)$ is not, we have shown that the orientation dependence of the dynamics initiated by a probe pulse can be obtained. Unlike in weak-field RCS, where perturbative excitation of the rotational wavepacket restricts the sum in Eq.~\ref{eq:fit} to $j_{max} = 2$, a strong-field rotational wavepacket contains high-order coherences~\cite{ren2013,spector2014} that can be used to extract the orientation dependence with correspondingly higher resolution.      

To the best of our knowledge,  Fig.~\ref{fig:ion2D} represents the first experimental determination of the two dimensional angle dependence of a non-dissociative process from a polyatomic molecule; the corresponding time-dependent molecular axis distribution (shown in SM) is also the first measurement of its kind. These represent important steps forward in gas phase molecular physics and chemistry as the principles and methods used to make these measurements are general and may be applied to any asymmetric top molecule, and apply to physical processes that can be described within the approximation that rotational and vibronic motion can be separated.  For instance, the 2D angle dependent amplitudes and phases of harmonics from HHG could serve as raw material for orbital tomography \cite{itatani2004,vozzi2011}, extraction of the fully differential photorecombination cross section \cite{bertrand2012,ren2013}, or the understanding of the electronic motion occurring occurring in the ion between the ionization and recombination steps leading to HHG \cite{smirnova2009, kraus2015}. Similar analysis of the time dependence of photoelecton angular distributions in the laboratory frame will provide molecular frame photoelectron angular distributions shown to be a useful probe of excited state molecular dynamics~\cite{underwood2008}. Thus, extending the work presented here to other processes can provide invaluable insights into complex systems.

\begin{acknowledgments}
This work was supported by the U.S. Department of Energy, Office of Science,
Basic Energy Sciences, under Award \#DE-FG02-86ER13491.
\end{acknowledgments}
\bibliography{references}

\end{document}